\documentclass[twocolumn,amsmath,amssymb,superscriptaddress]{revtex4}

\usepackage{graphicx}
\usepackage{dcolumn}
\usepackage{bm}

\begin{document}


\title{Exact solutions of Lovelock-Born-Infeld Black Holes}

\author{Mat\'\i as Aiello}
\email{aiello@iafe.uba.ar}
\affiliation{Departamento de F\'\i sica, Facultad de Ciencias Exactas y 
Naturales, Universidad de Buenos Aires, Ciudad Universitaria, Pabell\'on
I, 1428 Buenos Aires, Argentina}
\author{Rafael Ferraro}
\email{ferraro@iafe.uba.ar}
\thanks{Member of Carrera del Investigador Cient\'{\i}fico (CONICET,
Argentina)}
\affiliation{Departamento de F\'\i sica, Facultad de Ciencias Exactas y
Naturales, Universidad de Buenos Aires, Ciudad Universitaria, Pabell\'on
I, 1428 Buenos Aires, Argentina}
\affiliation{Instituto de  Astronom\'\i a y F\'\i sica del Espacio, 
Casilla de Correo 67, Sucursal 28, 1428 Buenos Aires, Argentina}
\author{Gast\'on Giribet}
\email{gaston@ias.edu}
\affiliation{Departamento de F\'\i sica, Facultad de Ciencias Exactas y
Naturales, Universidad de Buenos Aires, Ciudad Universitaria, Pabell\'on
I, 1428 Buenos Aires, Argentina}
\affiliation{Institute for Advanced Study, Einstein Drive, Princeton 
NJ08540}


\begin{abstract}
The exact five-dimensional charged black hole solution in Lovelock
gravity coupled to Born-Infeld electrodynamics is presented. This
solution interpolates between the Hoffmann black hole for the
Einstein-Born-Infeld theory and other solutions in the Lovelock
theory previously studied in the literature. The conical
singularity of the metric around the origin can be removed by a
proper choice of the black hole parameters. The thermodynamical
properties of the solution are also analyzed and, in particular,
it is shown that the behaviour of the specific heat indicates the
existence of a stability transition point in the vacuum solutions.
We discuss the similarities existing between this five-dimensional
geometry and the three-dimensional black hole. Like BTZ black
hole, the Lovelock black hole has an infinite lifetime.

\end{abstract}

\pacs{Valid PACS appear here}
\maketitle

\section{Introduction}
The Einstein tensor is the only symmetric and conserved tensor
depending on the metric and its derivatives up to the second
order, which is linear in the second derivatives of the metric.
Dropping the last condition, Lovelock \cite{lovelock} found the
most general tensor satisfying the other ones. The obtained tensor
is non linear in the Riemann tensor and differs from the Einstein
tensor only if the space-time has more than 4 dimensions.
Therefore the Lovelock theory is the most natural extension of
general relativity in higher dimensional space-times. The Lovelock
theory for a particular choice of the coefficients of the action
could be thought as the gravitational analogue of Born-Infeld
electrodynamics \cite{ban}.

In the last decades a renewed interest in both Lovelock gravity and
Born-Infeld electrodynamics has appeared because they emerge in
the low energy limit of string theory \cite{frad,berg,met}. Since
the Lovelock tensor contains derivatives of the metric of order
not higher than the second, the quantization of the linearized
Lovelock theory is free of ghosts. For this reason the Lovelock
Lagrangian appears in the low energy limit of string theory. In
particular, the Gauss-Bonnet terms (quadratic in the Riemann
tensor) were studied in Ref. \cite{zwi} and the quartic terms in
Refs. \cite{cai,gross}. The Lovelock theory of gravity was also
discussed in Refs. \cite{joel,cai2,izaurrieta}.

Hoffmann was the first one in relating general relativity and the
Born-Infeld electromagnetic field \cite{Hoffmann}. He obtained a
solution of the Einstein equations for a  point-like Born-Infeld
charge, which is devoid of the divergence of the metric at the
origin that characterizes the Reissner-Nordstr\"{o}m solution.
However, a conical singularity remained there, as it was later
objected by Einstein and Rosen. The Einstein-Born-Infeld black 
hole has been revisited in Refs. \cite{gibb,oli}

The aim of this paper is to study the charged black hole solutions
in five-dimensional Lovelock gravity coupled to Born-Infeld
electrodynamics. In section II, we discuss the five-dimensional
Lovelock gravity and study the black hole solutions as a
preliminary step toward the calculation of the charged solution
which we present in a separated section. In section III, we
discuss the Born-Infeld electrodynamics which will provide us the
necessary tools in order to eventually find, in section IV, the
five- dimensional charged black hole solution in
Lovelock-Born-Infeld field theory. We study the geometrical
properties of the solution and discuss the similarities and
distinctions existing with respect to the Reissner-Nordstr\"om
black hole. Section V is dedicated to the analysis of the
thermodynamics of the Lovelock black holes and the conclusions are
contained in section VI.

\section{Lovelock theory}
The Lovelock Lagrangian density in $D$ dimensions is \cite{lovelock,deru}
\begin{equation}
{\cal L}=\sum_{k=0}^{N} \alpha_{k} \: \lambda^{2(k-1) } \:
{\cal L}_k
\label{lov1}
\end{equation}
where $N= \frac{D}{2}-1$ (for even $D$) and $N=\frac{D-1}{2}$ (for
odd $D$). In (\ref{lov1}), $\alpha _k$ and $\lambda $ are
constants which represent the coupling of the terms in the whole
Lagrangian and give the proper dimensions.

In  Eq. (\ref{lov1}) ${\cal L}_k$ is
\begin{equation}
{\cal L}_k = \frac {1}{2^k} \sqrt{-g}\:
\delta^{i_1...i_{2k}}_{j_1...j_{2k}}\: R^{j_1 j_2}_{i_1
i_2}...\:R^{j_{2k-1}j_{2k}}_{i_{2k-1}i_{2k}} \label{lov2}
\end{equation}
where ${R^{\mu}\:_{\nu\rho\gamma}}$ is the Riemann tensor in $D$
dimensions,
$R^{\mu\nu}\:_{\rho\sigma}=g^{\nu\delta}\:R^{\mu}\:_{\delta\rho\sigma}$,
$g$ is the determinant of the metric $g_{\mu\nu}$ and
$\delta^{i_1...i_{2k}}_{j_1...j_{2k}}$ is the generalized
Kronecker delta of order $2k$ \cite{{grav}}.
\bigskip

The Lagrangians up to order 2 are given by
\cite{deru,lan,lan1}
\bigskip

${\cal L}_0 = \sqrt{-g}$

\bigskip

${\cal L}_1 = \frac{1}{2}\: \sqrt{-g} \: \delta^{i_1 i_2}_{j_1
j_2}\:R^{j_1 j_2}_{i_1 i_2}= \sqrt{-g} \: R\;\;\;$

\bigskip

${\cal L}_2 = \frac{1}{4}\: \sqrt{-g} \: \delta^{i_1 i_2 i_3
i_4}_{j_1 j_2 j_3 j_4}\: R^{j_1 j_2}_{i_1 i_2}\:R^{j_3 j_4}_{i_3
i_4} = \;\:\:$
\begin{displaymath}
\;\;=\sqrt{-g}\: (
R_{\mu\nu\rho\sigma}\:R^{\mu\nu\rho\sigma}-4\: R_{\mu\nu}\:
R^{\mu\nu}+R^2)
\end{displaymath}
where we recognize the usual Lagrangian for the cosmological term,
the Einstein-Hilbert Lagrangian and the Lanczos Lagrangian
\cite{lan,lan1} respectively. 

For dimensions $D=5$ and $D=6$ the Lovelock Lagrangian is a linear
combination of the Einstein-Hilbert and Lanczos Lagrangians. The
spaces of dimensions $D=7$ and $D=8$ includes the Lagrangian ${\cal
L}_3$, which  was first obtained by  M\"uller-Hoissen \cite{mul}.
In general, if the dimension is even and the manifold is compact
with positive definite metric, then  ${\cal L}_{D/2}$ is the Euler
characteristic classes generator \cite{pat}
\begin{displaymath}
\int_{\mho}\:d^{2n}x\:{\cal L}_n=\frac{(-1)^{n+1}\:
2^{2+n}\:\pi^n\: n!}{(2n)!}\chi(\mho)
\end{displaymath}
where $\chi(\mho)$ are the Euler-Poincar\'e topological
invariants, which are zero in odd dimensions.

Hence, the geometric action is written as
\begin{equation}
S=\int d^Dx\:{\cal L}
\label{accion}
\end{equation}
From the variational principle we obtain the Lovelock tensor
(${\cal G}_{\mu\nu}$),
\begin{equation}
\delta S = \int d^Dx\:\delta{\cal L} = \int d^Dx \:{\cal
G}_{\mu\nu}\:\sqrt{-g}\:\delta g^{\mu\nu}
\end{equation}
where the general expression of ${\cal G}_{\mu\nu}$ is
\cite{lovelock,lovelock1}
\begin{equation}
{\cal G}^\mu \: _\nu(k)= -\frac{1}{2^{k+1}}\:\delta_{\nu\: j_1
j_2....j_{2k}}^{\mu\: i_1 i_2....i_{2k}}\:R^{j_1 j_2}_{i_1
i_2}\:....R^{j_{2k-1} j_{2k}}_{i_{2k-1} i_{2k}}
\end{equation}
\begin{equation}
{\cal G}_{\mu\nu}=\sum_{k=0}^{N} \alpha_{k}
\:\lambda^{2(k-1)}\:{\cal G}_{\mu\nu}(k)
\label{tensorl}
\end{equation}
where $N=\frac{D}{2}-1$ for even $D$ or $N=\frac{D-1}{2}$ for odd $D$. When 
the dimension is $D=4$, ${\cal G}_{\mu\nu}$ differs from the
Einstein tensor in a divergence term. The Lovelock tensors for
$k=0$ and $k=1$ are
\begin{displaymath}
{\cal G}_{\mu\nu}(0)=g_{\mu\nu}
\end{displaymath}
\begin{displaymath}
{\cal G}_{\mu\nu}(1)=R_{\mu\nu}-\frac{1}{2}\:g_{\mu\nu}\:R
\end{displaymath}

In five dimensions the Lagrangian is a linear combination of the
Einstein-Hilbert and the Lanczos ones, and the Lovelock tensor
results
\begin{displaymath}
{\cal G}_{\mu\nu}=R_{\mu\nu}-\frac{1}{2}\:R\:g_{\mu\nu}+\Lambda\:g_{\mu\nu}
\end{displaymath}
\begin{displaymath}
-\alpha\:\{\frac{1}{2}\:g_{\mu\nu}\:(R_{\rho\delta\gamma\lambda}\:R^{\rho
\delta\gamma\lambda}-4\:R_{\rho\delta}\:R^{\rho\delta}+R^2)-
\end{displaymath}
\begin{equation}
2\:R\:R_{\mu\nu}+4\:R_{\mu\rho}\:R^{\rho}_{\nu}+4\:R_{\rho\delta}
\:R^{\rho \delta}_{\mu \nu}-2\:R_{\mu\rho\delta\gamma}\:R_{\nu}^{\rho\delta\gamma}\}
\label{love}
\end{equation}

To be precise, let us notice that in the third term of Eq. (\ref{love}) we are abusing of the notation 
when including the cosmological constant $\Lambda$ explicitely. We do 
that with the intention to indicate the nature of each term in a clear 
way,
even though the constant $\Lambda$ is actually related to the couplings
$\alpha_0$ and $\lambda$ in Eq. (\ref{tensorl}) by the relation $\Lambda =
-\frac{\alpha_0}{2\lambda ^2}$.
The Gauss-Bonnet constant $\alpha=\alpha_2 \lambda ^2$ will allow us to 
track the
changes in the equations, when we compare with the respective ones
of general relativity. The coupling constant $\alpha $ introduces
a length scale $l_L\sim \sqrt {\alpha }$ in the theory which
physically represents a short-distance range where the Einstein
gravity turns out to be corrected.

The vacuum field equations are given by
\begin{equation}
{\cal G}_{\mu\nu}=0 \;\; ; \; 0\:\le\: \mu\:,\:\nu\: \le 4
\label{ecuvacio}
\end{equation}
and accept spherically symmetric solutions in five dimensions,
which in terms of a suitable Schwarzschild-like {\it ansatz}, can
be written as
\begin{displaymath}
ds^2\:=\:-\:\Psi(r)\:dt^2+\: \frac{1}{\Psi(r)}\:dr^2+\:r^2\:d
\theta^2+\:r^2\:sin^2 \theta\:d\chi^2
\end{displaymath}
\begin{equation}
+\:r^2\:sin^2 \theta\: sin^2 \chi \:d\varphi^2
\label{intervalo}
\end{equation}
In this case, the solution of Eqs. (\ref{ecuvacio}) is
\begin{equation}
\Psi^{\pm}(r)= 1+\frac{r^2}{4\:\alpha}\pm \sqrt{1+\frac{M}{6\:\alpha}+
\frac{r^4}{16\:\alpha^2}+\frac{\Lambda}{12\:\alpha}\:r^4}
\label{sol}
\end{equation}
where  $M$ is an integration constant.

By requesting the proper Newtonian potential in the weak field
region $r \rightarrow \infty$ ($\Lambda=0$), it results that the
ADM mass is $m=\frac{\pi}{6}M+\pi\:\alpha$ \cite{louko} with $\alpha>0$ 
for $\Psi^-$ and $\alpha<0$ for $\Psi^+$. Then
\begin{equation}
\Psi(r)=1+\frac{r^2}{4\:\alpha}-\frac{r^2}{4\:\alpha}\:\sqrt{1
+\frac{16\:m\:\alpha}{\pi\:r^4}+\frac{4\:\alpha\:\Lambda}{3}}
\label{solnueva2}
\end{equation}
Asymptotically, this solution goes to the general relativity
solution in five dimensions when $\alpha \to 0$, as it is
expected. Namely
\begin{equation}
\Psi_{GR}(r)=1-\frac{2\:m}{\pi\:r^2}-\frac{\Lambda}{6}\:r^2
\label{sol1}
\end{equation}

Let us notice that in the case of non-vanishing cosmological constant, besides the leading term in the expansion (\ref{sol1}), we find finite-$\alpha $ corrections to the black hole parameters. Namely

\begin{equation}
\Psi (r) = 1-\frac {2m_{d}}{\pi r^2}-\frac {\Lambda _d}{6} r^2 +{\cal O} (\alpha r^{-6})
\end{equation} 
where the {\it dressed} parameters $m_d$ and $\Lambda _d$ are given by
\begin{eqnarray*}
\Lambda _d &=& \Lambda \left( 1+ \sum _{n=2}^{\infty } c_n \mu ^{n-1 }\right) \\ 
m _d &=& m \left( 1+ \sum _{n=2}^{\infty} n \ c_n \mu ^{n-1} \right) 
\end{eqnarray*} 
being 
\begin{equation*}
c_n = \frac{(2n-3)!!}{2^{n-1}n!} \ , \ \ \ \mu = -\frac {4}{3} \Lambda \alpha . 
\end{equation*}
Furthermore, in the case of the charged solution we will discuss in section IV, the (charge) parameter $Q$ receives similar corrections due to these finite-$\alpha $ effects, resulting
\begin{equation*}
Q^2_{d}= Q^2 \left( 1+ \sum _{n=2}^{\infty} n \ c_n \mu ^{n-1} \right) 
\end{equation*}
Notice that the parameter $\mu $ controls the {\it dressing} of the whole set of black hole parameters. The above power expansion converges for values such that $\mu < 1$. Besides, for the case $\mu > 1$ we find a different expansion, leading to the following {\it dressed} parameters in the large $r$ regime
\begin{equation*}
m_{d} = \frac {m}{\sqrt{| \mu | }} \left( 1+\sum _{n=2}^{\infty} n \ c_n 
\mu ^{1-n}
\right)
\end{equation*}
Thus, we note that the Newtonian term $\sim m_d r^{-2}$ vanishes in the limit
$| \Lambda \alpha | \to \infty$. The particular case $\mu = 1$ is discussed
below. Moreover, it is possible to see that, if one considers the case $\alpha\Lambda > 0$, the effective cosmological constant in the large $\mu $ limit turns out to be 
\begin{equation*}
\Lambda _d = \sqrt{\frac{3\Lambda}{\alpha}} -\frac {3}{2\alpha }+ {\cal O} (1/\sqrt {|\mu |}) \ .
\end{equation*}

On the other hand, for the case of vanishing cosmological constant ($\Lambda=0$), the
solution (\ref{solnueva2}) displays an event horizon located at
$r_h=\sqrt{\frac{2\:m}{\pi}-2\:\alpha}$ when  $m \geq \pi\:\alpha$. Then,
the horizon can reach the point $r=0$ for a massive object with
$m=\pi\:\alpha$; in this case $r=0$ is a naked singularity. If
$m<\pi\:\alpha$ there is no horizon.

One of the relevant differences existing between the black hole
solutions in Einstein and Lovelock theories is the fact that
$\Psi(r)$ is not singular at the origin. Instead, the metric is
regular everywhere. From (\ref{sol}) we obtain
\begin{displaymath}
\Psi(r=0)=1-\sqrt{\frac{m}{\pi\:\alpha}}
\end{displaymath}
and, in fact, we will find a similar aspect for the case of the
Lovelock-Born-Infeld charged black hole.

If the object has no mass ($m=0$), one gets de Sitter ($a<0
\Rightarrow \Lambda>0$) and anti-de Sitter (AdS) solutions ($a>0
\Rightarrow \Lambda<0$) as particular cases; namely
\begin{displaymath}
\displaystyle \Psi_{(A)dS}(r)=1+a\:r^2
\end{displaymath}
where $\displaystyle
a=\frac{1}{4\:\alpha}\:\bigg(1-\sqrt{1+\frac{4}{3}\:\Lambda\:\alpha}\bigg)$

Another interesting geometry is found in the particular case
$\alpha \Lambda = -\frac 34$. At this point of the space of
parameters, the solution (\ref{sol}) becomes
\begin{equation}
\Psi_{BTZ}(r) = \frac{r^2}{4\:\alpha}-{\cal M}
\label{gaston}
\end{equation}
where we have considered $\Lambda <0$, and introduced the notation
${\cal M} +1= \sqrt{\frac{m}{\pi\:\alpha}} $. Certainly, we could
refer to this particular black hole solution as the {\it BTZ
branch}, due to its reminiscence of BTZ black hole
\cite{btz,btz2}; this is an aspect that was already pointed out in Ref. \cite{banados}. Actually, the solution (\ref{gaston}) shares
several properties with the three-dimensional black hole geometry,
as the thermodynamical properties which will discussed in section V.
Indeed, the parameter ${\cal M}$ in Eq. (\ref{gaston}) plays the
role of the mass $M_{BTZ}$ in the BTZ solution. For instance, as
well as $AdS_3$ spacetime is obtained as a particular case of the
BTZ geometry by setting the negative mass $M_{BTZ}=-1$, also the
five-dimensional Anti-de Sitter space corresponds to setting
${\cal M} = -1$.

On the other hand, let us notice that, in a consistent way, if the
large $\alpha $ limit is taken while fixing the condition $\Lambda
\alpha =-\frac 34$ one finds that the solution becomes the metric
which represents the near boundary limit of $AdS_5$, like it
happens with the {\it massless BTZ} ($M_{BTZ}=0$) which is
obtained by making the three-dimensional black hole to disappear.
Then, the parallelism with the solutions in $D=3$ turns out to be
exact since the five-dimensional metric obtained by keeping only
the leading terms in the near boundary limit of $AdS_5$
corresponds to ${\cal M} =0$ in (\ref{gaston}) as well, which is
precisely the Lovelock black hole solution (\ref{sol}) with {\it
minimal} mass $m= \pi\alpha $. Besides, a conical singularity is
found in the range $0<m<\pi\alpha $ (corresponding to $-1 <{\cal
M}<0$) in a complete analogy.

The $AdS$-symmetry invariance of this particular Lovelock solution
was also discussed in Ref. \cite{banados}. Here, we have shown how the
solution (\ref{gaston}) appears as a particular case of the
geometry (\ref{sol}).

The digression above relies in the fact that, besides the
parameter $m$, which classifies the black hole geometry, we have
additional parameters characterizing the field theory by means of
the couplings $\alpha _k$ of the different contributions ${\cal L}
_k$ in the Lovelock Lagrangian ${\cal L}$. This is the case of the
cosmological constant $\Lambda $ and the Gauss-Bonnet constant
$\alpha $. As an example, we observed that if one constrains the
theory on a particular curve in the space of parameters defined by
$\Lambda \alpha =-\frac 34$ one finds the particular solution
(\ref{gaston}). Thus, different regions of the space of parameters
($\Lambda , \alpha $) lead to quite different qualitative
behaviours of the theory and, consequently, of its solutions.

\section{Born-Infeld Electrodynamics}
In 1934 Born and Infeld \cite{born1,born} proposed a non-linear
electrodynamics with the aim of obtaining a finite value for the
self-energy of a point-like charge. The Born-Infeld Lagrangian
leads to field equations whose spherically symmetric static
solution gives a finite value $b$ for the electrostatic field at
the origin. The constant $b$ appears in the Born-Infeld Lagrangian
as a new universal constant. Following Einstein, Born and Infeld
considered the metric tensor $g_{\mu\nu}$ and the electromagnetic
field tensor $F_{\mu\nu}=\partial
_{\mu}A_{\nu}-\partial_{\nu}A_{\mu}$ as the symmetric and
anti-symmetric parts of a unique field $b\:g_{\mu\nu}+F_{\mu\nu}$.
Then they postulated the Lagrangian density
\begin{equation}
{\cal L}=\sqrt{\det (b\:g_{\mu\nu}+F_{\mu\nu})}+\sqrt{-\det
g_{\mu\nu}}
\label{BIL}
\end{equation}
where the second term is chosen so that the Born-Infeld Lagrangian
tends to the Maxwell Lagrangian when $b\rightarrow \infty$. In
four dimensions, this Lagrangian results to be
\begin{equation}
{\cal
L}=\sqrt{-g}\frac{b^2}{4\:\pi}\bigg(1-\sqrt{1+\frac{2S}{b^2}-
\frac{P^2}{b^4}}\bigg)
\end{equation}
where $S$ and $P$ are the scalar and pseudoscalar field invariants
\begin{displaymath}
S=\frac{1}{4}F_{\mu\nu}F^{\mu\nu}=\frac{1}{2}(B^2-E^2)
\end{displaymath}
\begin{displaymath}
P=\frac{1}{8}\sqrt{-g}\:\epsilon_{\mu\nu\rho\gamma}F^{\mu\nu}F^{\rho\gamma}=
E \cdot B
\end{displaymath}

The Born-Infeld Lagrangian is usually mentioned as an exceptional
Lagrangian because its properties of being the unique structural
function which: 1- Assures that the theory has a single
characteristic surface equation; 2- Fulfills the positive energy
density and the non-space like energy current character
conditions; 3- Fulfills the strong correspondence principle. As a
consequence of these conditions, the Lagrangian has time-like or
null characteristic surfaces \cite{pleb}.

In order to obtain the static spherically symmetric solution in
five dimensions, we will replace $F=E(r)\:dt\wedge dr$ and the
metric (\ref{intervalo}) in the Born-Infeld Lagragian (\ref{BIL});
then we will vary the action (this procedure is valid due to the
high symmetry of the solution we are looking for). Therefore
\begin{equation}
{\cal L}_{BI}=\frac{b^2\:\sqrt{-g}}{4\:\pi}\Bigg(1-\sqrt{1-
\:\frac{E^2(r)}{b^2}}\Bigg) \label{lagrang}
\end{equation}
The field equation derived from this Lagrangian (\ref{lagrang}) is
\bigskip
\begin{displaymath}
\frac{\partial}{\partial r}\Bigg[ \frac{\sqrt{-g}\: E(r)
}{\sqrt{1-\:\frac{E^2(r)}{b^2}}}\Bigg]=0
\end{displaymath}
\bigskip
where $\sqrt{-g}=r^3\:\sin^2\theta \: \sin \chi$. So the
Born-Infeld point charge field in five dimensions is
\begin{equation}
E(r)=\frac{Q}{\sqrt{r^6+L^6}}\;\; ;  \;\;L=\bigg(\frac{Q}{b}\bigg)^{\frac{1}{3}}
\end{equation}
The energy-momentum tensor is
\begin{displaymath}
T_{\mu\nu}=-\frac{2}{\sqrt{-g}}\:\frac{\partial {\cal L}_{BI}}{\partial g^{\mu\nu}}
\end{displaymath}
\bigskip
In the static isotropic case it results to be diagonal:
\begin{displaymath}
T_0^0=T_r^r=\frac{b^2}{4\:\pi}\Bigg[\:1-\frac{1}{\sqrt{1-\:\frac{E^2(r)}{b^2}}}\Bigg]
\end{displaymath}
\begin{equation}
T_\chi^\chi=T_\theta^\theta=T_\varphi^\varphi=\frac{b^2}{4\:\pi}
\Bigg[\:1-\sqrt{1-\:\frac{E^2(r)}{b^2}}\Bigg] \label{energia}
\end{equation}

The energy of this field is finite in contrast to the energy of
the Maxwell field:
\begin{equation}
U=2\pi^2\int_{0}^{\infty} T_{0}^{0}\ r^3\ dr
=\frac{-b^2\:L^4\:\sqrt{\pi}}{12}\:\Gamma
\bigg(-\frac{2}{3}\bigg)\:\Gamma\bigg(\frac{1}{4}\bigg)
\end{equation}

\section{Lovelock-Born-Infeld solutions}
We will study the exact solutions of Lovelock gravity for a
Born-Infeld isotropic electrostatic source. The field equations to
be solved are ${\cal G}_{\mu\nu}=8\:\pi\:T_{\mu\nu}$, where ${\cal
G}_{\mu\nu}$ is the Lovelock tensor (\ref{love}) and $T_{\mu\nu}$
is the Born-Infeld energy-momentum tensor corresponding to a point
charge located in the origin (\ref{energia}). Because of the
symmetry of the source we repit the {\it ansatz} (\ref{intervalo})
for the metric. In this case only the diagonal components of the
Lovelock tensor survive: the components $\mu=\nu=0, 1$ are equal,
and they are integrals of the components $\mu=\nu=2, 3, 4$.
Therefore, it is enough to  solve ${\cal G}_{00}=8\:\pi\:T_{00}$,
which amounts to the equation
\bigskip
\begin{displaymath}
-\frac{\Psi(r)}{2\:r^3}\:\Bigg(3\:r^2\:\frac{d\Psi(r)}{dr}+12\:\alpha
\:\frac{d\Psi(r)}{dr}-12\:\alpha\:\Psi(r)\:\frac{d\Psi(r)}{dr}+6\:\Psi(r)\:r
\end{displaymath}
\begin{displaymath}
-6\:r+2\:r^3\:\Lambda\Bigg) =\frac{2\:b^2\:\Psi(r)}{r^3}\:(\sqrt{r^6+L^6}-r^3)
\end{displaymath}
The left hand side can be written as a total radial derivative, to
be easily integrated. The solution is
\begin{displaymath}
\Psi^{\pm}(r)=1+\frac{r^2}{4\:\alpha}\pm
\Bigg(1+\frac{M}{6\:\alpha}+\bigg(\frac{1}{4\:\alpha}-\frac{4
\:b^2-2\:\Lambda}{6}\bigg)\:\frac{r^4}{4\:\alpha}
\end{displaymath}
\begin{equation}
+\frac{2\:b^2}{3\:\alpha}\:\int^r_0\:dr\:\sqrt{r^6+L^6}\Bigg)^{\frac{1}{2}}
\label{soll}
\end{equation}
where $M$ is an integration constant. The integral inside the
square involves an incomplete elliptic integral of the first kind
$F(a,b)$ \cite{franklin}, namely
\bigskip
$\displaystyle
\int^r_0\:dr\:\sqrt{r^6+L^6}=\frac{1}{4}\:r\:\sqrt{r^6+l^6}+\frac{3}{4}L^4\:
\int^{\frac{r}{L}}_0\:\frac{dt}{\sqrt{1+t^6}}$, where
\bigskip
$\displaystyle \int^{\frac{r}{L}}_0\:\frac{dt}{\sqrt{1+t^6}}=$

\bigskip

$\displaystyle \frac{1}{2\:3^{\frac{1}{4}}}\:F\Bigg(\arccos \Bigg[\frac{L^2+(1-\sqrt{3})\:r^2}{L^2+(1+\sqrt{3})\:r^2}\Bigg],\frac{2+\sqrt{3}}{4}\Bigg)$

\bigskip
Thus, we obtain two solutions for the metric, but the sign of
$\alpha$ is determined requiring that in the limit $r \rightarrow
\infty$ we must recover the Newtonian potential in five dimensions 
$\Phi=\frac{m}{\pi r^2}$, so we obtain
$\alpha>0$ for $\Psi^-$ and $\alpha<0$ for $\Psi^+$. In that limit
the solution is

\bigskip
$\displaystyle \Psi(r)=1-\frac{2\:m}{\pi\:r^2}$ with $\displaystyle 
m=\frac{\pi}{6}M+\pi\:\alpha+\frac{\pi}{2}\:L^4\:b^2\:\gamma\;\;$

\bigskip
where  $\displaystyle \gamma= \int_0^\infty \frac{dt}{\sqrt{1+t^6}}=\frac{\Gamma\bigg(\frac{1}{3}\bigg)\:\Gamma\bigg(\frac{7}{6}\bigg)}{\sqrt{\pi}}\approx 1.40218$.

\bigskip
In terms of the ADM mass $m$, $\Psi(r)$ becomes
\begin{displaymath}
\Psi(r)=1+\frac{r^2}{4\:\alpha}-\frac{r^2}{4\:\alpha}\:\Bigg(1+\frac{16\:m\:\alpha}{\pi\:r^4}-\frac{2}{3}\:\alpha\:(4\:b^2-2\:\Lambda)
\end{displaymath}
\begin{equation}
+\frac{8\:b^2\:\alpha}{3\:r^3}\:\sqrt{r^6+L^6}-\frac{8\:b^2\:L^6\:\alpha}{r^4}\:\int_{r}^{\infty} \frac{dr}{\sqrt{r^6+L^6}}\Bigg)^{\frac{1}{2}}
\label{solteo}
\end{equation}
This class of solutions was also studied in reference \cite{wiltshire2}.

In the limit $\alpha \rightarrow 0$ the solution tends to

\begin{displaymath}
\Psi(r)\approx 1-\frac{\frac{6}{\pi}m-2\:L^4\:b^2\:\gamma}{3\:r^2}-\frac{\Lambda}{6}\:r^2-
\end{displaymath}
\begin{equation}
\frac{4\:b^2}{3\:r^2}\:\int^r_0\:dr\:\bigg(\sqrt{r^6+L^6}-r^3\bigg)
\label{limitealfa}
\end{equation}
This limit agrees with the quoted four-dimensional Hoffmann
solution \cite{Hoffmann} with a conical singularity in the origin
of the black hole. Consequently, by performing the limit $b
\rightarrow \infty$ in Eq. (\ref{limitealfa}) we recover the
Reissner-Nordstr\"om solution in five dimensions with cosmological
constant,
\begin{displaymath}
\Psi_{RN}(r)=1-\frac{2\:m}{\pi\:r^2}+\frac{Q^2}{3\:r^4}-\frac{\Lambda}{6}\:r^2
\end{displaymath}

On the other hand, by taking the limit $b \rightarrow \infty$ in
(\ref{solteo}), we also recover the Lovelock-Maxwell solution
found by Wiltshire in \cite{w}; namely
\begin{equation}
\Psi_{W}(r)=1+\frac{r^2}{4\alpha}-\frac{r^2}{4\alpha} \bigg [1+\frac{16\:m\alpha}{\pi\:r^4}
-\frac{8Q^2\alpha}{3r^6}+\frac{4\Lambda \alpha}{3}\bigg]^{\frac{1}{2}}
\end{equation}
Notice that, because to the existence of a Birkhoff-like theorem
(see Appendix A in Ref. \cite{louko} and references therein), this
limit turns out to be more than a simple heuristical argument to
check the solution (\ref{solteo}), representing a necessary
condition which required to be proved. We also observe that the uncharged solution (11) is recovered in the limit $b \to 0$.

\begin{figure}
\begin{center}
\includegraphics[width=7cm,angle=0]{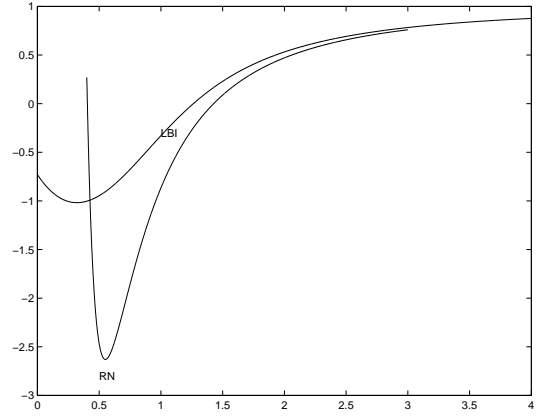}
\caption[]{Comparison between Reisnner-Nordstr\"{o}m (RN) and $\Psi(r)$ (LBI) [$\alpha=0.1$, $Q=1$, $\Lambda=0$, $b=1$, $m=\pi$].}
\label{comparacion}
\end{center}
\end{figure}

Coming back to the general solution (\ref{solteo}), we can see
that, differing from Schwarzschild and Reissner-Nordstr\"{o}m
solutions, the metric has no singularities (Fig.
\ref{comparacion})
\begin{equation}
\Psi(r=0)=1-\bigg(\frac{m}{\pi\:\alpha}-\frac{b^2\:L^4\:\gamma}
{2\:\alpha}\bigg)^{\frac{1}{2}}
\label{cerosdepsi}
\end{equation}

Depending on the values of the parameters of the black hole
(charge $Q$, mass $m$, Born-Infeld constant $b$ and Gauss-Bonnet
constant $\alpha$) the square root could be imaginary. If $\alpha
>0$ and  $m\geq \frac{\pi}{2}\:b^2\:L^4\:\gamma=m_c\;\;$ the solution is
valid for all $r\geq 0$.

If $m>m_c$ then $\Psi(0) \neq 1$, so the metric has a conical
singularity at the origin: whereas the equator measures
$2\:\pi\:r$, the radius at the equator is
$\int_o^r\:\frac{dr}{\sqrt {\Psi(r)}}$. In the critical case
$m=m_c $ one finds $ \Psi(0)=1$ and the conical singularity
disappears from the metric since in this case $
\int_o^r\:\frac{dr}{\sqrt {\Psi(r)}}\approx r$ when $r \approx 0$.

Conversely, we can think in the critical value $m_c$ as follows: we can define a critical value $Q^2 _c = b^{-1}\left( \frac {2m}{\pi \gamma} \right)^{3/2}$ which represents an upper bound on $Q$ for the metric (24) to be well defined in the whole spacetime. Thus, a critical value for the black hole charge appears in this context as a direct consequence of the finite-$b$ effects. The role payed by $b$ is setting the critical value $Q_c \to 0 $ in the Maxwellian regime $b\to \infty $, which is consistent with the fact that the Lovelock-Maxwell black hole geometry is not regular if $m \neq 0$.

Besides, the finite-$b$ effects act in (\ref{solteo}) as a kind of effective cosmological constant $\Lambda _{eff} (r) = \Lambda +2b^2(1-e^{u_{b}(r)})$, with $u_0 = 0$ and $\displaystyle \lim _{r \to \infty } u_b (r) =0$. Thus, this fact could lead one to infer that the {\it dressing} of the black hole parameters discussed in section II can also receive contribution due to the presence of $b$. However, this is not the case, as it can be verified by noting that no $b$-dependent quadratic terms in $r$ arise when expanding the right hand side of (\ref{solteo}). 

\begin{figure}
\begin{center}
\includegraphics[width=7cm,angle=0]{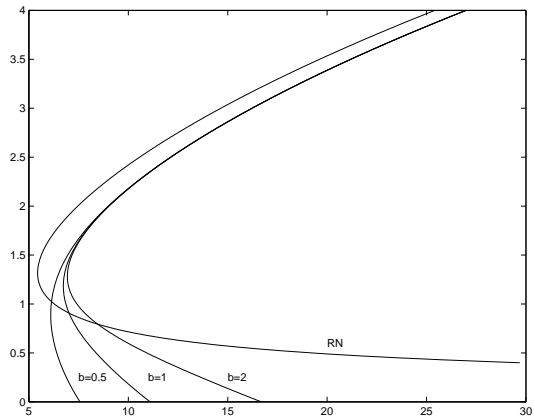}
\caption[]{Event horizon position (vertical axis) as a function of the
black hole mass (horizontal axis) for
different values of $b$ and in the Reissner-Nordstr\"{o}m
case (RN) [$\alpha=0.5$, $Q=3$, $\Lambda=0$].}
\label{mvsh}
\end{center}
\end{figure}

\begin{figure}
\begin{center}
\includegraphics[width=7cm,angle=0]{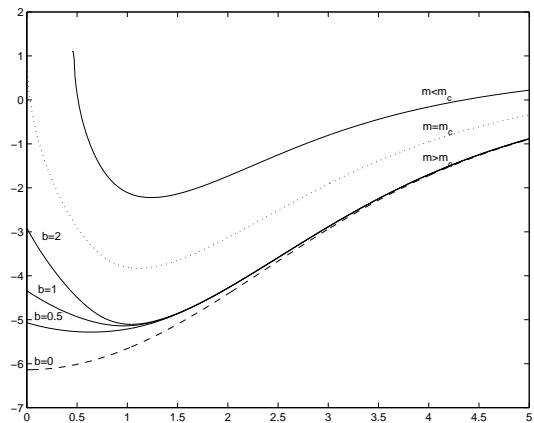}
\caption[]{Graphics of $\Psi(r)$ for different values of $b$, for $m<m_c$: 
[$m=2$, $\alpha=0.5$, $Q=1$, $b=2$, $\Lambda=0$], for $m=m_c$ 
[$m=3.49$, $\alpha=0.5$, $Q=1$, $b=2$, $\Lambda=0$] and for $m>m_c$ 
[$m=5$, 
$\alpha=0.5$, $Q=1$, $\Lambda=0$].}
\label{b1}
\end{center}
\end{figure}

In this geometry, the position of the horizon $r_h$ is defined by $\Psi(r_h)=0$,
then
\begin{displaymath}
r_h^2=\frac{2\:m}{\pi}-2\:\alpha-\frac{r_h^4}{12} 
(4\:b^2-2\:\Lambda)
+\frac{b^2}{3}\:r_h\:\sqrt{r_h^6+L^6}
\end{displaymath}
\begin{displaymath}
-b^2\:L^4\:\int_{r_h / L}^{\infty}\:\frac{dt}{\sqrt{1+t^6}}
\end{displaymath}
and thus, for $\Lambda=0$ and $m>m_c+\pi\:\alpha$ there is only one horizon.
If $m \le m_c+\pi\:\alpha$ then the solution is similar to
Reissner-Nordstr\"{o}m in the sense that there could be two
horizons. When the equality holds one of the horizons is at the
origin (see Eq. (\ref{cerosdepsi})). Figure \ref{mvsh} shows the
position of the horizon ($r_h$) as a function of the mass $m$ for
the Lovelock-Born-Infeld black hole and the Reissner-Nordstr\"{o}m
case ($b \rightarrow \infty$).
These graphics exhibit the crucial difference existing between the
Lovelock-Born-Infeld black hole and the Reissner-Nordstr\"om black hole; namely the fact that for a given charge $Q$
there exists a finite value for the black hole mass $m$ such that
the black hole geometry presents only one horizon, because the
{\it internal} one reached the origin. This enhancement of the
region bounded by both internal and external horizons is also
$b$-dependent and represents, by itself, one of the principal
distinctions between the black hole geometries in both theories.
Moreover, Fig. \ref{mvsh} shows how the extremal configuration
$r_+=r_-$, which is translated into a complicated expression in
terms of the parameters $m,Q,\alpha$ and $b$, experiments a
displacement for finite values of $b$ with respect to the
Reissner-Nordstr\"om configuration $m^2=\frac{\pi^2}{3}Q^2$.
Figure \ref{b1} shows different behaviors of the solution $\Psi(r)$ for 
different values of b. This parameter controls the cualitative behaviour 
close to the origin. The metric for the subcritical case ($m<m_c$) is 
not 
well defined in the
whole spacetime.
  
\section{Thermodynamics}
\begin{figure}
\begin{center}
\includegraphics[width=7cm,angle=0]{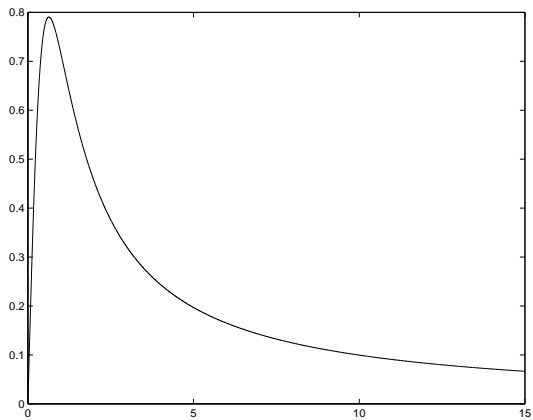}
\caption[]{Hawking temperature as a function of the horizon radius [$\alpha=0.1$, $\Lambda=0$].}
\label{hawtem}
\end{center}
\end{figure}

Now, let us move to the thermodynamics of the vacuum solution. The
Hawking temperature of the black hole is proportional to the
surface gravity $\kappa$ and it is given by the formula
\begin{equation}
T=\frac{\hbar}{4 \pi k_B}\kappa
\label{temperat}
\end{equation}
where $k_B$ is the Boltzmann constant. In order to calculate
$\kappa$ we must evaluate the derivative of $g_{00}$ in the
exterior horizon radius $r_h$. From the Lovelock vacuum solution
(\ref{solnueva2}) with $\Lambda=0$ we obtain
$r_h=\sqrt{\frac{2\:m}{\pi}-2\alpha}$. Hence, the surface gravity of
the corresponding geometry results
\begin{displaymath}
\kappa=\frac{\sqrt{\frac{2m}{\pi}-2\alpha}}{\alpha(1+\frac{m}{\pi\alpha})}
=\frac{2\:r_h}{4\alpha+r_h^2}
\end{displaymath}
and the Hawking temperature is given by
\begin{equation}
T=\frac{\hbar}{2\pi k_B} \frac{r_h}{4\alpha+r_h^2}
\label{temp1}
\end{equation}
Notice that the temperature goes to zero for bigger black holes
($r_h \rightarrow \infty$). This behaviour is the same of the
Schwarzschild black hole. A remarkable results is that, unlike the
Schwarzschild case, for small black holes ($r_h \rightarrow 0$ or
$m \rightarrow m_c$) the temperature also goes to zero and there
would not be Hawking radiation (Fig. \ref{hawtem}). On the other
hand, when $\alpha \rightarrow 0$ in equation (\ref{temp1}) we
recover the general relativity result since, in that case, the
temperature is proportional to the inverse of the radius.

\begin{figure}
\begin{center}
\includegraphics[width=7cm,angle=0]{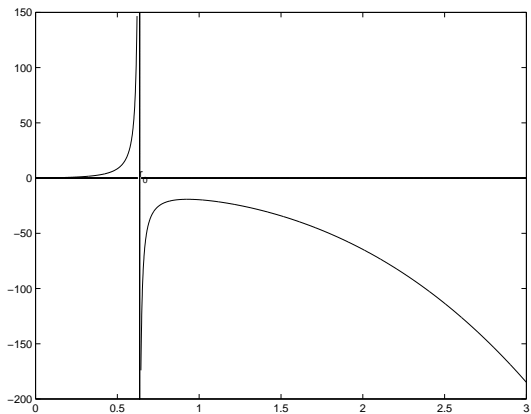}
\caption[]{Black hole specific heat as a function of the horizon radius [$\alpha=0.1$, $\Lambda=0$].}
\label{caloresp}
\end{center}
\end{figure}

The specific heat $c$ is given by
\begin{equation}
c=\frac{\partial m}{\partial r_h}\Bigg( \frac{\partial T}{\partial r_h}
\Bigg)^{-1}=\frac {2\pi^2 k_B}{\hbar 
}\frac{r_h(4\alpha+r_h^2)^2}{4\alpha-r_h^2}
\label{calores}
\end{equation}
and, from this, we observe that the position $r_0=\sqrt{4\alpha}$
$(m=3 \pi \alpha )$, where the specific heat changes the sign (Fig.
\ref{caloresp}), represents a transition point (in the Fig.
\ref{hawtem} this point is the maximum). When  $r_h<r_0$  the
specific heat is positive and the black hole is stable; while in
the range $r_h>r_0$ the specific heat is negative and the black
hole turns out to be, in this sense, unstable. The temperature of
this transition point is $\displaystyle T_0=\frac{\hbar}{2 \pi
k_B}\frac{1}{\sqrt{4\alpha}}$.

Regarding the entropy of this black hole solution, we make use of
the fact that the first law for this non-rotating black hole can
be written as usual,
\begin{equation}
T\:\delta S=\delta m
\label{tempdeh}
\end{equation}
where $\delta m=\pi\:r_h\:\delta r_h$. Thus the entropy results
\begin{equation}
S=\frac{ \pi^2 k_B}{2 \hbar}(12\alpha\:r_h+r_h^3)
\label{entrdeh}
\end{equation}

In the limit $\alpha \rightarrow 0$ the entropy is proportional to
the area of the event horizon, in agreement with the
Bekenstein-Hawking formula. On the other hand, the first term in
the equation (\ref{entrdeh}) is a consequence of the Gauss-Bonnet
term. It is possible to see that this correction term in the
entropy formula is proportional to the integral of the scalar
curvature invariant of the horizon \cite{wald}. This computation
confirms, by means of a concise example, the observation derived
from Hamiltonian methods in \cite{mayer} about the fact that the
entropy of Lovelock black holes is not simply the area formula,
but topological invariants also contributes to the whole entropy. This aspect was also discussed in \cite{myers2}.

Besides, by using the Stefan-Boltzmann law in five dimensions to
compute the flux of thermal radiation we can approximately obtain
the black hole evaporation time $\Delta t$; namely
\begin{equation}
\frac{dm}{dt}=\pi r_h \frac{dr_h}{dt} \propto T^5 r_h^3
\label{tiempo}
\end{equation}

With the expression for the Hawking temperature (\ref{temp1}) and 
integrating (\ref{tiempo}) one finds that the
lifetime of the Lovelock black holes turns out to be
infinite. This is due to the linear dependence between the
temperature and the horizon radius for small black holes. Hence, these are eternal black
holes.

\begin{figure}
\begin{center}
\includegraphics[width=7cm,angle=0]{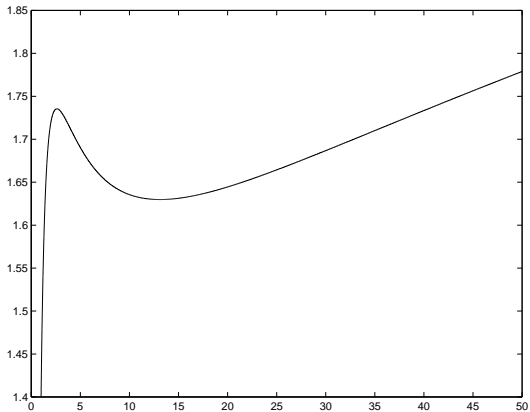}
\caption[]{Hawking temperature for a negative cosmological constant as a function of the mass [$\alpha=0.1$, $\Lambda=-0.6$].}
\label{temp6}
\end{center}
\end{figure}

Furthermore, we find that the themodynamics of the Lovelock black
holes in presence of a negative cosmological constant is also
consistent with the known results of Einstein gravity solutions
within the appropriate range. Certainly, the Lanczos Lagrangian
represents short-distance corrections to Einstein gravity which
start to be relevant at scales comparable with $l_L \sim \sqrt
{\alpha }$, which we could denominate {\it Lanczos-Lovelock
scale}. Conversely, the value of the cosmological constant
$\Lambda $ introduces another length scale, given by $l_{\Lambda
}\sim \Lambda ^{-2}$, which marks the scales where the
cosmological term starts to dominate. Hence, we can identify the
range bounded by the maximum and the minimum of Fig. \ref{temp6}
as the scales where the results coming from general relativity fit
the thermodynamic behaviour of the Lovelock solution.
Consequently, for very large scales (large black holes) the
thermodynamic behaviour is well approximated by the
AdS-Schwarzschild black hole solution while the short distance
corrections start to dominate at the scale $l_L$.

\begin{figure}
\begin{center}
\includegraphics[width=7cm,angle=0]{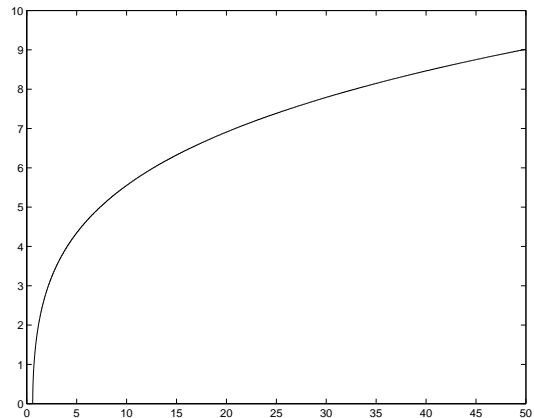}
\caption[]{Hawking temperature for the case $\Lambda \alpha=-\frac{3}{4}$ as a function of the mass [$\alpha=0.1$, $\Lambda=-7.5$].}
\label{temp34}
\end{center}
\end{figure}

Within this context, we can also notice that the thermodynamical
properties of the black hole geometry that appears in the case $\Lambda
\alpha =-\frac 34$ (with $\Lambda <0$) exhibit particular features.
Indeed, as we early mentioned, this case is analogous to the BTZ
three-dimensional black hole since the phase diagram involving the
temperature and the mass is a monotonic function (see Fig. \ref{temp34})
describing a phase such that the specific heat is positive everywhere. In
some sense, this case can be considered as a transition and, thus, its
stability could be an interesting point to be studied with particular
attention. The temperature of these black holes is given by $T=\frac{\hbar
}{2\pi k_B} \frac{r_h}{4\alpha}$; this goes to zero for large $\alpha $
and identically vanishes for the critical mass $m=\pi\alpha $.

\section{Conclusions}
In this paper we studied a solution (\ref{solteo}) representing
five-dimensional charged black holes in Lovelock gravity coupled
to Born-Infeld electrodynamics. The corrections induced by the
quadratic terms in the Lagrangian (Gauss-Bonnet terms) correspond
to short-distance modifications to general relativity and,
therefore, the relevant differences between both theories appear
for small radius.

The Lovelock-Born-Infeld black holes are characterized by the mass
($m$), the Gauss-Bonnet constant ($\alpha$), the charge ($Q$) and
the Born-Infeld constant ($b$). The constant $\alpha$ must be
positive in order to have  a well behaved solution for all value of $r$.
The metric does not diverge at $r=0$; for the critical mass
$m=m_c$ the conical singularity, which is characteristic of the
Hoffmann-Born-Infeld solution, is removed (nevertheless, the
origin is a curvature singularity).

We commented the differences existing with respect to the
Reissner-Nordstr\"om black holes and, from this analysis, we
observed that, unlike the general relativity, the
Lovelock-Born-Infeld theory admits charged black hole solutions
with only one horizon. This is due to the fact that for a given
charge $Q$, there exist values of mass that force the internal
black hole radius to reach the origin.

There is another important distinction between the solutions of
both theories. Also in contrast to the Schwarzschild solution, the
temperature of the black hole remains finite; in particular, we
showed that the temperature goes to zero when the horizon radius
approximates to the origin, and there is not Hawking radiation.
This leads to find an infinite lifetime for Lovelock solutions
because the short-distance effects render the small black holes
stable. The temperature of the transition point, where the
short-distance corrections start to be relevant, is
$T=\frac{\hbar}{\pi k_B}\frac{1}{\sqrt{4\alpha}}$, which
corresponds to black holes with a size comparable to the
Lovelock-Lanczos scale, $r=2 l_L$.

We discussed different limits of the solutions in terms of the
coupling constant of Lanczos Lagrangian $\alpha $ and Born-Infeld
Lagrangian $b$, and we proved that these limits correspond to the
expected geometries. Hence, the solution we present here
represents a geometry interpolating between the quoted Hoffmann
metric for Einstein-Born-Infeld theory and the solution found by
Wiltshire for the case of Lovelock-Maxwell field theory.
Furthermore, we showed how other solutions studied in the
literature are included as particular cases, representing a {\it
BTZ phase} which arises on the curve $\Lambda \alpha = -\frac 34$
in the space of parameters. We discussed the similar features of
this phase and the Anti-de Sitter black holes.

\appendix

\acknowledgments

R.F. was supported by Universidad de Buenos Aires (UBACYT X103) and 
Consejo Nacional de Investigaciones Cient\'\i ficas y T\'ecnicas 
(Argentina).

G.G. was supported by Fundaci\'on Antorchas and Institute for 
Advanced 
Study; on leave of absence from Universidad de Buenos Aires.

We thank M. Kleban, J.M. Maldacena, R. Rabad\'an, R. Troncoso and J.
Zanelli for useful discussions.

\end{document}